\documentclass[article,twocolumn,showpacs,superscriptaddress]{revtex4-1}

\usepackage{graphicx}
\graphicspath{{./},{figs/}}

\usepackage{amssymb}
\usepackage{bm}
\usepackage{braket}
\usepackage{amsmath}
\usepackage{xcolor}
\usepackage[normalem]{ulem}

\usepackage[colorlinks=true,
                linkcolor=blue,
	        citecolor=blue,
	        urlcolor=blue]{hyperref}

\newcommand\etal{{\it et al}.\ }

\newcommand{\cnrs}{Laboratoire Charles Coulomb (L2C), Université de Montpellier, CNRS, Montpellier, France}
\newcommand{\dqmp}{Department of Quantum Matter Physics, University of Geneva, CH-1211 Geneva, Switzerland}
\newcommand{\unimore}{Dipartimento di Scienze Fisiche, Informatiche e Matematiche, University of Modena and Reggio Emilia, IT-41125 Modena, Italy}
\newcommand{\cnr}{Centro S3, CNR-NANO Istituto Nanoscienze, IT-41125, Modena, Italy}
\newcommand{\argonne}{Materials Science Division, Argonne National Laboratory, Lemont, Illinois 60439, USA}
\newcommand{\chicago}{Department of Physics, University of Chicago, Chicago, Illinois 60637, USA}

\AtBeginDocument{\usepackage{booktabs}}
\begin{document}

\title{Unconventional gate-induced superconductivity in  transition-metal dichalcogenides}

\date{\today}

\author{Thibault Sohier}
\affiliation{\cnrs}

\author{Marco Gibertini}
\affiliation{\unimore}
\affiliation{\cnr}

\author{Ivar Martin}
\affiliation{\argonne}
\affiliation{\chicago}

\author{Alberto F.\ Morpurgo}
\affiliation{\dqmp}

\begin{abstract}
Superconductivity in few-layer semiconducting transition metal dichalcogenides (TMDs) can be induced by field-effect doping through ionic-liquid gating. While several experimental observations have been collected over the years, a fully-consistent  theoretical picture is still missing. Here we develop a realistic framework that combines the predictive power of first-principles simulations with the versatility and insight of Bardeen-Cooper-Schrieffer gap equations to rationalize such experiments. The multi-valley nature of semiconducting TMDs is taken into account, together with the doping- and momentum-dependent electron-phonon and Coulomb interactions. 
Consistently with experiments, we find that superconductivity occurs when the electron density is large enough that the Q valleys get occupied, as a result of a large enhancement of electron-phonon interactions.
Despite being phonon-driven, the superconducting state is predicted to be sensitive to Coulomb interactions, which can lead to the appearance of a relative sign difference between valleys and thus to a $s_{+-}$ character. 
We discuss qualitatively how such scenario may account for many of the observed physical phenomena for which no microscopic explanation has been found so far, including in particular the presence of a large subgap density of states, and the sample-dependent dome-shaped dependence of $T_c$ on accumulated electron density.
Our results provide a comprehensive analysis of gate-induced superconductivity in semiconducting TMDs, and introduce an approach that will likely be valuable for other multivalley electronic systems, in which superconductivity occurs at relatively low electron density.
\end{abstract}

\maketitle


\section{Introduction}

Gate-induced superconductivity at the surface of semiconducting transition metal dichalchogenides (TMDs) has been discovered 10 years ago, in pioneering experiments on ionic gated transistors based on exfoliated MoS$_2$ crystals \cite{Ye2012,Taniguchi2012}. Subsequent work \cite{Shi2015,Jo2015,Lu2015,Lu2018} showed that superconductivity occurs in the majority of the commonly studied semiconducting TMDs (WS$_2$, MoSe$_2$), starting at different, but comparable, values of accumulated electron density. The maximum critical temperature  in the different  compounds is also comparable, and reaches up to  $T_c \simeq 11$ K in MoS$_2$ at 10$^{14}$ electrons/cm$^{-2}$~\cite{Ye2012,Costanzo2016}.  The occurrence of gate-induced superconductivity in semiconducting TMDs is therefore a robust phenomenon, that manifests itself in a similar way in the majority of compounds in this class~\cite{Qiu2021}, including atomically thin samples~\cite{Costanzo2016,Fu2017,Lu2018,Zheliuk2019,Ding2022}. Despite its experimental robustness, the  nature of the gate-induced superconducting state remains currently unknown.

A number of facts have been established experimentally. The in-plane critical magnetic field, for instance, significantly exceeds the Pauli limit~\cite{Lu2015,Saito2016}, a consequence of the very strong spin-orbit interaction present in semiconducting TMDs~\cite{Zhu2011,Xiao2012,Manzeli2017} that causes the spin of the electrons forming Cooper pairs to point in the direction normal to the layers. This implies that superconductivity is of the Ising type~\cite{Zhou2016}, and that electron pairs have to form in a superposition of singlet and triplet states. It has also been established that the onset of the gate-induced superconducting state coincides with the filling of electronic states in the Q-valleys~\cite{Piatti2018,Piatti2019}. 
Interestingly, in thick layers of common semiconducting TMDs, the conduction band edge at the Q-point is located at an energy lower than that of the band-edge at the K-point \cite{Roldan2014,Yuan2016}, but in transistor devices the large perpendicular electric field that accompanies electron accumulation deforms the conduction band. This deformation causes the band-edge at the K-point to be shifted to lower energy than that at the Q-point already at rather low electron density (typically, lower than  $\simeq 10^{13}$ cm$^{-2}$), and experiments~\cite{Piatti2018} correlating the evolution of normal state transport with the occurrence of superconductivity indicate that gate-induced superconductivity first occurs when both the K- and Q-valley are populated  (note that intercalated MoS$_2$ bilayer devices show superconductivity of  different nature already at lower density, with only the K/K' and valley occupied). It has also been concluded from gate-dependent Raman spectroscopy experiments~\cite{Sohier2019} that a very large enhancement of the electron-phonon coupling strength takes place in transistors of all the investigated semiconducting TMDs when the Q-valley starts to be populated by electrons. Experiments therefore strongly suggest that the transition to the gate-induced superconducting state is  driven by electron-phonon coupling.

Additional experiments providing useful indications --albeit with less clear-cut microscopic implications-- have also been reported. For instance, the critical temperature of the gate-induced superconducting state appears to exhibit a dome-like shape as a function of accumulated electron density $n$, reaching a maximum at an optimal value of $n$, before decreasing for larger electron accumulation. Details, however, seem to depend on the material and on the specific experiment reported. In some cases the ``dome'' is unambiguously present (e.g., in MoS$_2$~\cite{Ye2012}) whereas in others it is less pronounced, i.e., it is less clear whether $T_c$ actually decreases or just tends to saturate (e.g., for the highest mobility WS$_2$ monolayers~\cite{Lu2018,Ding2022}). Systematic tunnelling conductance measurements have been performed on  MoS$_2$ transistors and have shown a large sub-gap density of states, with the zero-bias tunneling conductance scaling linearly with temperature down to $T\ll T_c$~\cite{Costanzo2018}. Such a behavior is incompatible with the presence of a fully gapped superconducting state, and suggests that gate-induced superconductivity has unconventional character. Whether this behavior is common to all TMDs is however not yet established, as controlled tunneling conductance measurements  have so far been performed exclusively on MoS$_2$.

Our goal here  is  to investigate theoretically the nature of the superconducting state by combining first principles calculations with Bardeen-Cooper-Schrieffer (BCS) theory~\cite{Bardeen1957}, to rationalize the experimental observations. Semiconducting TMDs are particularly well-suited to implement such a strategy, because a large body of work has already been done to analyze their normal state properties in terms of ab-initio calculations~\cite{Li2007,Kadantsev2012,Kormanyos2015,Brumme2015}, and to compare results to experiments~\cite{Splendiani2010,Wang2021}. This past work has established that all important aspects of the low-energy band structure of common semiconducting TMDs, its dependence on perpendicular electric field, and details of the electron-phonon and Coulomb interactions in the presence of accumulated electron density are correctly captured by first principles calculations, with an overall excellent quantitative agreement. It should therefore be expected that ab-initio simulations  also allow capturing the key ingredients needed to describe the gate-induced superconducting state, namely the multivalley structure of these materials, the strong and strongly momentum dependent electron-phonon interaction, and Coulomb repulsion. 

Taking into account these ingredients at a sufficient level of detail is important because most earlier attempts to describe gate-induced superconductivity in semiconducting TMDs were based on too drastic approximations. Examples include early work that largely overestimated $T_c$ by only considering  the effect of electron-phonon interaction~\cite{Ge2013,Rosner2014}, or phenomenological models that included exclusively Coulomb repulsion to analyze the possible occurrence of unconventional superconductivity~\cite{Roldan2013,Yuan2014}. Typically, this early work only considered states in the K and K' valley, which --as we now understand-- is not appropriate to explain experimental results. Over the years,  more refined models--even generalized to hole-doping~\cite{Hsu2017,Oiwa2018}--have been introduced, for instance including spin and density fluctuations~\cite{Das2015}, employing quantum Monte Carlo~\cite{Wang2022}  or using a valley-dependent $\mu$-parameter to include Coulomb repulsion effects~\cite{Schonhoff2016}, but still with limited agreement with experiments. Very recently, a first-principles approach~\cite{Margine2013} based on a momentum resolved Migdal-Eliashberg theory~\cite{Migdal1958,Eliashberg1960} and including anharmonic effects~\cite{Monacelli2021} has been applied to the case of TMDs~\cite{Marini2023}, but with a treatment of Coulomb repulsion that misses the crucial role of non-local interactions in 2D~\cite{Simonato2023} and multivalley effects~\cite{Das2023}.

Focusing on monolayers, we formulate a realistic model of multi-valley effects within the simplest BCS approach, by employing ab-initio techniques to analyze the momentum-resolved electron-phonon and Coulomb interaction as a function of accumulated electron density. The results of this analysis are used to determine the interaction coupling constants for intra-valley and inter-valley processes responsible for electron pairing, which are subsequently  included in the BCS equation for the critical temperature. Such an approach represents  a sensible approximation at the  electron densities reached by electrostatic gating, which are much smaller than those of common metallic superconductors (and which therefore result in relatively small Fermi surfaces). It is particularly effective, as it allows important physical effects to be captured while keeping the formalism simple, which greatly facilitates their interpretation. 

We find that the superconducting transition is driven by the large enhancement of the electron-phonon interaction strength that occurs when the Q-valleys start to be populated, in agreement with both existing Raman spectroscopy experiments~\cite{Sohier2019} (which directly probe electron-phonon interaction) and $T_c$ measurements~\cite{Ye2012,Taniguchi2012,Shi2015,Jo2015,Lu2015,Lu2018,Costanzo2016, Piatti2018} (the value of $T_c$ that we obtain are comparable to the experimental ones). We further find that --even though superconductivity is driven by electron-phonon interaction-- Coulomb repulsion can profoundly affect  the nature of the superconducting state, as it can cause a sign change in some of the intervalley interaction coupling constants. When that happens, the relative sign of the order parameter in different valleys become opposite and the superconducting state acquires an $s_{+-}$ character. Our model therefore supports an unusual scenario, in which $s_{+-}$ superconductivity occurs despite the  dominant interaction driving superconductivity being electron-phonon attraction. We discuss qualitatively how --in actual devices-- such  scenario may account for some of the observed physical phenomena for which no microscopic explanation has been found so far, such as the presence of a large subgap density of states, and the sample-dependent dome-shaped dependence of $T_c$ on accumulated electron density. 

The paper is structured as follows. In Sec.~\ref{sec:model} we first introduce the key quantities and ingredients entering the theoretical model and their evaluation from first principles. The model predictions concerning  gate-induced multivalley superconductivity in semiconducting TMD monolayers are then presented in Sec.~\ref{sec:results}. Finally, in Sec.~\ref{sec:final} we summarize the main findings and discuss their implications towards the interpretation of existing experiments.

\section{Model}\label{sec:model}

This work focuses on semiconducting TMD single layers.
In this section, we first describe the rich electronic structure of those monolayers and then propose a BCS-like model for superconductivity adapted to their multivalley character. This model is able to capture how the interplay between  electron-phonon attraction and Coulomb repulsion determines the pairing within and between valleys.
It is then reported how those two competing interactions are computed from first-principles simulations, as a function of momentum and doping (i.e.\ Fermi level).

\subsection{Multivalley structure of TMDs}

\begin{figure}[h]
    \centering
    \includegraphics[width=\linewidth]{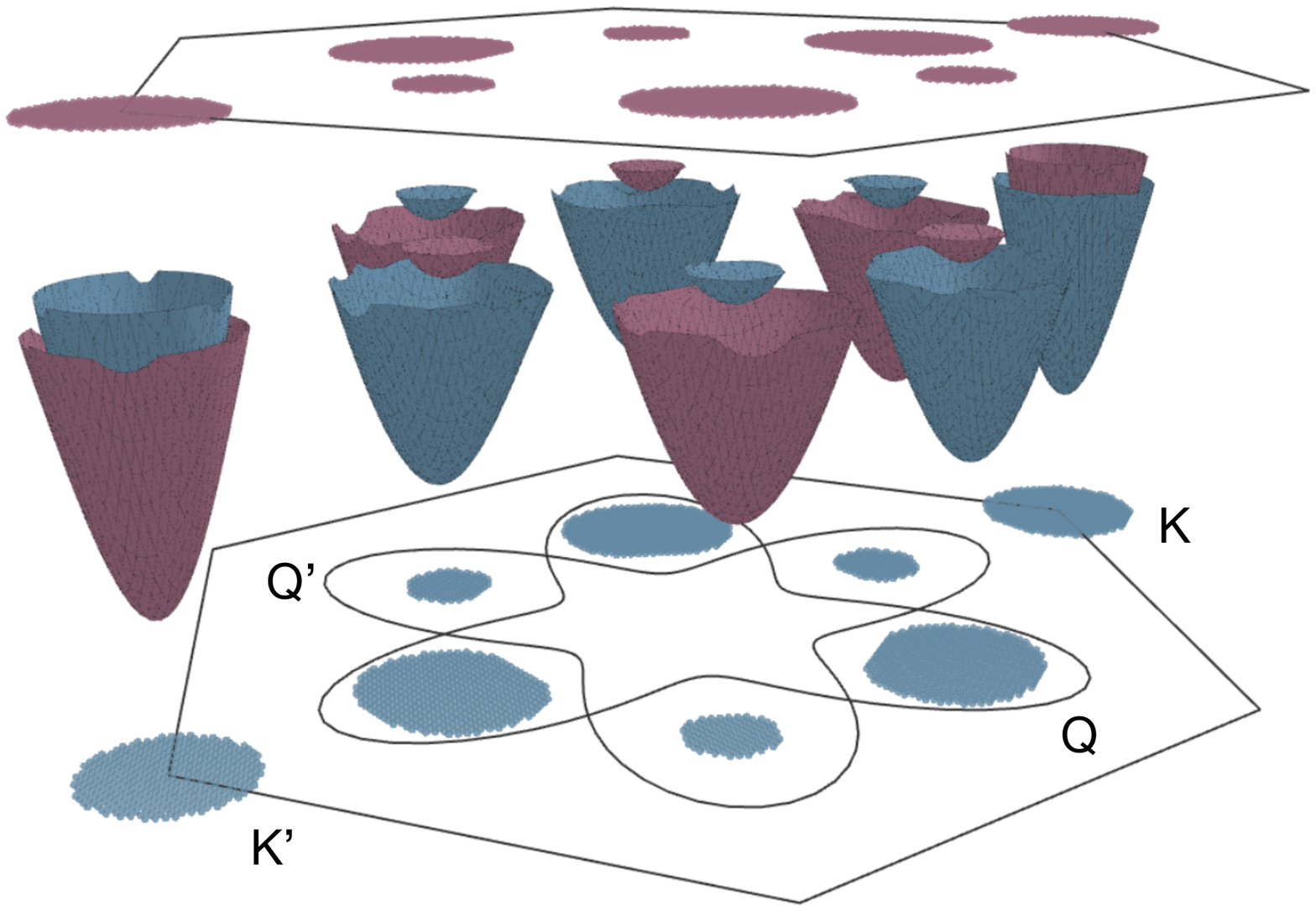}
    \caption{Spin and valley structure of WS$_2$. Opposite spins are projected on opposite sides of the valleys. We assume independent spin channels and consider only one spin (blue) in the model. We group the 3 Q (or Q') valleys that are degenerate in energy. Note that only two non-equivalent K points are represented, such that all states of the first BZ are represented once and only once.}
    \label{fig:SpinValleyStructure}
\end{figure}

Semiconducting TMD monolayers display a rich electronic structure, with multiple valleys~\cite{Brumme2015} and a distinct spin texture~\cite{Xiao2012,Xu2014}, both in the valence and in the conduction bands. 
In the following we shall consider mainly electron doping, and we thus focus on the lowest conduction states that are occupied by electrons upon gating. 
In Fig.~\ref{fig:SpinValleyStructure} we show the corresponding bottoms of the  conduction bands across the hexagonal Brillouin zone (BZ), as computed in the case of WS$_2$ by density-functional-theory (DFT) simulations (see Sec.~\ref{sec:methods} for more details). 
Multiple valleys are present, whose structure is further enriched by a sizable spin-orbit coupling that, combined with the lack of inversion symmetry, gives rise to a net spin splitting. The spin polarization is essentially orthogonal to the layers, with spin up and spin down states represented respectively in blue and red in Fig.~\ref{fig:SpinValleyStructure}. 
Although results are illustrated with the specific case of WS$_2$, the  valley and spin structure is quite general and applies to all the TMDs considered here (MoS$_2$, MoSe$_2$, WS$_2$, and WSe$_2$), with only the quantitative details for the energy separation between different valleys and different spins varying~\cite{Roldan2014,Kormanyos2015}. 

The lowest energy conduction states are located around two high-symmetry points at the opposite BZ vertices, typically denoted as K and K', giving rise to two inquivalent valleys (the other BZ vertices being equivalent by a reciprocal lattice vector translation). The spin splitting in these valleys is small and opposite at K and K' because of time-reversal symmetry. At K/K', the splitting has different signs for W-based and Mo-based TMDs, but spin-split bands cross in Mo-TMDs relatively close to the band edge because they have different effective masses, resulting in the same spin ordering of the bands sufficiently far from the BZ vertices in the energy range relevant for this study. Additional valleys appear higher in energy, approximately located midway between the BZ vertices and the BZ center $\Gamma$, around six inequivalent k-points typically denoted with the letter Q (or less frequently $\Lambda$). For these valleys the spin splitting is much larger and alternates  in sign in a way that is consistent with the three-fold symmetry and time-reversal invariance. Overall, we thus have four distinct band edges, corresponding to the K and Q points for the two spin polarizations, giving rise to a total of 16 pockets, as shown in Fig.~\ref{fig:SpinValleyStructure}: $4=2\times2$ arising from spin-up and down states around K and K' and $12=2\times6$ associated with opposite spins at the six Q valleys. 

Given that Coulomb interaction does not flip spin and spin-conserving electron-phonon interactions are much stronger than the spin-flipping ones even in the presence of spin-orbit coupling in TMDs~\cite{Wang2018}, the spins of individual electrons comprising Cooper pairs are assumed to be conserved during scattering. The spin-orbit interaction in these materials is of Ising type, with an out-of-plane spin-quantization axis. Opposite spin pairing is expected to be favored energetically due to the time-reversal symmetry, which ensures that there is an equal energy partner with an opposite spin and momentum available for each electron (Note that this pairing is neither singlet nor triplet, but rather a linear combination of them~\cite{Gorkov2001}. In this context,  spin-up and spin-down channels can be considered to be equivalent and independent. In the following we shall therefore focus on a single channel and we  identify Cooper pairs  by just one of the partners, that we take to be spin up, i.e.\ we consider the blue valleys in Fig.~\ref{fig:SpinValleyStructure}. 

For what concerns the BZ corners, we thus have a valley at K and one at K', with the same spin but different energies. The six Q valleys with spin up can be separated into two groups that lie at  different energies, as shown in Fig.~\ref{fig:SpinValleyStructure}. The valleys within each group are related by the three-fold rotation symmetry, which enforces their degeneracy. In the following we shall consider superconducting solutions where this rotation symmetry is preserved, so it is convenient to pool together valleys within a given group into a single set. The lowest three valleys, which form a triangle that points towards the K' point, will be called Q, while the upper three, pointing towards the K point, will be denoted Q' (the choice of prime notation will become clear when discussing electron-phonon interactions). The superconducting gap equations will be written assuming that the electronic states belong to one of these four valleys: K, K', Q, and Q'.

The relevance of these four valleys depends on the doping level, which can be efficiently tuned in semiconducting TMDs using electrostatic gating~\cite{Zhang2019,Piatti2019,Ahn2006,Cao2023}. Mimicking experiments, the doping density (and Fermi energy) will thus be the main external parameter that we tune in our simulations. At low doping, only the K/K' valleys are occupied. With increasing density also the Q valleys start to get populated. The precise value for the onset of Q occupation with density does not affect qualitatively the present analysis and we do not claim to be predictive as this onset depends on the energy difference between the K and Q valleys, which is not known precisely at a quantitative level~\cite{Sohier2023} (large variations in the values are reported in both experiments and DFT simulations). At even larger density, the Q' valleys could also be occupied. However, the energy difference between Q and Q’ is generally large and it is unlikely that electrons visit the Q' valleys in existing experiments, i.e., it is unlikely that the Q' valley can be populated at the maximum density accumulated by ionic gating. Q' states are nevertheless kept in the model for completeness.

\subsection{Basic aspects of multiband SC and microscopic interactions}

We now describe our approach to modeling superconductivity. The Ising spin-orbit interaction guarantees that for every electron with momentum-spin $({\bf k}, \sigma)$ there is also an electron with the same energy at $(-{\bf k}, -\sigma)$. These electrons can form zero-momentum Cooper pairs via a weak coupling instability, in a way directly analogous to the standard BSC superconductivity. What makes the situation nonstandard in TMDs is that there are multiple Fermi surfaces--controlled  by the doping level selected by the voltage applied to the ionic gate--such that   the superconducting order parameter may vary both within each one of them and even more among them. Given that these Fermi surfaces are rather small relative to the Brillouin zone size,  we can use a simplified description where each Fermi pocket has a unique value of the order parameter, $\Delta_v$ ($v$ is the valley index). 

Different $\Delta_v$ are connected to each other by the self-consistent BCS  gap equations. These equations generalize the standard BCS equations for a single band \cite{Tinkham} and have the form~\cite{Das2023,Horhold2023}
\begin{align}
\Delta_v = M_{vw}\Delta_w.
\label{eq:eigen}
\end{align}
In general, the matrix $M$ depends both on temperature and the order parameters $\Delta_v$ themselves, $M = M(T, \Delta_v)$. Setting temperature to zero leads to a nonlinear system of equations for the order parameter in the ground state; setting $\Delta_v $ to zero in $M$ yields a linearized  equation in the order parameters that allows to compute the superconducting transition temperature, $\det [M (T_c)-\openone] = 0$.

\begin{figure}[h]
    \centering
    \includegraphics[width=0.49\textwidth]{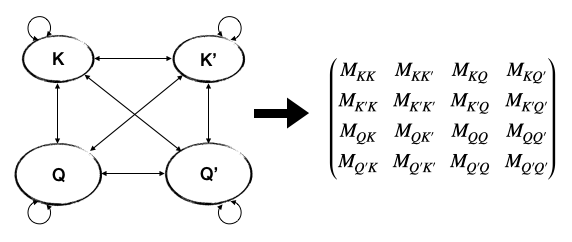}
    \caption{Multivalley superconductivity is modeled via a matrix containing the average interactions between electrons in different valleys.}
    \label{fig:MPSCmodel}
\end{figure}

Graphically, the matrix $M$ can be represented as depicted in Fig.~\ref{fig:MPSCmodel}.
Individual coupling matrix elements include, for a pair of valleys $vw$, the competition between phonon-mediated attraction ($G_{vw}$) and Coulomb repulsion ($C_{vw}$). 
\begin{align}
\label{eq:matrix}
    M_{vw} &= \ln{\left[
\frac{\omega_{vw}^{\rm eff}}{k_bT_c}\right]} \big(G_{vw} - C_{vw}\big)  \nonumber \\
    G_{vw} &= \left\langle \sum_{\mathbf{k}' \in w} \sum_{\nu} \frac{g^2_{\mathbf{k}\mathbf{k}', \nu}}{\omega_{\mathbf{q}, \nu}}   \delta(\varepsilon' - \varepsilon_F) \right\rangle_{\mathbf{k}\in v} \\
    C_{vw} &= \alpha \times\left\langle  \sum_{\mathbf{k}'\in w} V(q = |\mathbf{k}'-\mathbf{k}|) \delta(\varepsilon' - \varepsilon_F)\right\rangle_{\mathbf{k}\in v} \nonumber
\end{align}
with $v,w \in \{K, K', Q,Q'\}$. $M_{vw}$ represents an average matrix element for a Cooper pair scattering between valleys $v$ and $w$, all near the Fermi level $\varepsilon_F$.
A constituent electron of momentum $\mathbf{k}$ and energy $\varepsilon$ scatters to state $\mathbf{k}'=\mathbf{k}+\mathbf{q}$ ($\varepsilon'$) either via phonons or Coulomb interaction. $\langle f(\mathbf{k}) \rangle_{\mathbf{k}\in v}$ stands for the average of $f$ over $\mathbf{k}$ momenta in $v$ at $\varepsilon_F$. However, very minimal dependence on the choice of $\mathbf{k} \in v$ was found, as shown in methods, due to the fairly small Fermi surfaces. In practice, one reference state $\mathbf{k} \in v$ is picked from the $\varepsilon = \varepsilon_F$ iso-energetic line.
The matrix $M$ is computed for each material  at various carrier concentrations by evaluating from first-principles the phonon and the Coulomb components of $M$ as described in the Sect.~\ref{subsec:phonon} and \ref{subsec:coulomb}. 
In particular, we compute the electron-phonon matrix element $g_{\mathbf{k}\mathbf{k}', \nu}$, which describes the coupling between a phonon of mode $\nu$ at momentum $\mathbf{q}$ and frequency $\omega_{\mathbf{q},\nu}$ with two electronic states at $\mathbf{k}$ and $\mathbf{k}'=\mathbf{k}+\mathbf{q}$.
To account for Coulomb scattering, we compute the screened Coulomb potential within the isotropic approximation,
\begin{align}
\label{eq:scrCoul}
    V(q) &= \frac{2\pi e^2}{q \epsilon(q)},  
\end{align}
where $q =|\mathbf{q}|$ and $\epsilon(q)$ is the dielectric function in the static approximation.  It includes screening by itinerant electrons (See Sect.~\ref{subsec:coulomb} for details), and offsets the attractive interaction provided by phonons. Within the BCS treatment, it thus comes in with the same energy range of integration and therefore the same logarithmic prefactor as the attraction $G$.

There is a possibility of a more accurate Eliashberg-type treatment~\cite{Eliashberg1960} that allows for fully frequency dependent order parameter, up to the electronic band-width frequencies, which can be adapted to a first-principles approach~\cite{Margine2013,Lucrezi2024}. In that case, the bare Coulomb interaction is taken as unscreened and the screening is generated automatically in the course of solving the frequency depended Eliashberg equations for the normal and anomalous Greens functions and self-energies \cite{Morel1962}. Our approach is more qualitative and simple: we assume that the dominant pairing is due to phonons, which defines the relevant frequency window around the Fermi level, while the Coulomb interaction is taken as electronically screened. For the latter we take advantage of the first principles methods as well; this allows us to be material-specific not only as far as the electron phonon interaction is concerned, but also for the Coulomb (pseudo)potential.
The limitation of our approach is that  is only justifiable if phonon-mediated attraction is the dominant pairing mechanism.

To uncover the role of Coulomb interaction, we introduce a scaling factor $\alpha$. It allows us to gradually turn on the Coulomb repulsion and explore how the nature of superconductivity changes with the interaction strength. In practice, the scaling factor $\alpha$  can approximately capture the dependence of Coulomb on the dielectric properties of the surrounding substrate/ionic liquid.

Each $M_{vw}$ has its own effective phonon (``Debye") frequency under the logarithm, which corresponds to the energy of the typical phonons at the momenta involved in the inter- or intra-valley transitions. In practice we average over the phonon frequencies weighted by their coupling to the electrons:
\begin{align}
    \omega_{vw}^{\rm eff} &= \frac{\sum_{\mathbf{k}' \in w, \mathbf{k}\in v} \sum_{\nu} \int \frac{g^2_{\mathbf{k}\mathbf{k}',\nu}}{\omega} \ \omega  \ \delta(\varepsilon' - \varepsilon_F) }{\sum_{\mathbf{k}' \in w, \mathbf{k}\in v}\sum_{\nu}\int \frac{g^2_{\mathbf{k}\mathbf{k}',\nu}}{\omega} \delta(\varepsilon' - \varepsilon_F)} 
\end{align}

The superconducting transition is signaled by the appearance of a non-zero solution in the {\em linearized} (i.e. $\Delta\to0$) system of equations; it corresponds to the temperature for which the matrix $M$ develops a unit eigenvalue. 
The corresponding eigenvector represents the relative signs and sizes of the order parameter in  the different valleys. For a purely attractive interaction, by Perron-Frobenius theorem, all components $\Delta_v$ of the eigenvector with the largest eigenvalue have the same sign. This corresponds to the standard $s$-wave state, adjusted for the fact that there is Ising spin orbit coupling.
However, the elements of $M$ can change sign for some valley pairs if the Coulomb repulsion exceeds the phonon-mediated attraction.  
In this case, the largest eigenvalue (the highest transition temperature) can have an eigenvector with both positive and negative elements, implying an order parameter that has different signs on different Fermi sheets. This would indicate the formation of a $s_\pm$ state.

The added value of the approach described above is that it provides a good compromise between complexity and accuracy.
Electron-phonon interactions are highly dependent on both the magnitudes and the relative angles between momenta. 
In relatively crude estimations of superconducting properties, these interactions are averaged over all  Fermi surfaces and combined into a single $\lambda$ parameter~\cite{Ge2013,Zeng2016,Fu2017}. The absence of valley resolution  then limits the physical insight into the multivalley aspect of the coupling mechanisms.
In more sophisticated estimations, the full momentum dependency is retained within the Eliashberg framework~\cite{Marini2023}
(see also \cite{Heil2017,Paudyal2020} in other families of TMDs). While quantitatively more complete, the size and complexity of the resulting momentum-dependent order parameter makes it difficult to draw qualitative conclusions. Also, the role of different phonons can become obscured.
Here, we retain only the principal momentum structure of the couplings in the superconducting problem by averaging over valleys. This level of coarse-graining is found to be simple enough to provide physical insight, and detailed enough  to understand the underlying multivalley mechanisms. Furthermore, we  have the full momentum and energy dependence of the couplings 
and  can identify the phonons leading to dominant couplings within a given valley pair.
Concerning the Coulomb repulsion, most works simply use a constant parameter~\cite{Rosner2014,Zeng2016}, $\mu$. This is not satisfactory, as already demonstrated by earlier works that do account for the variation of $\mu$ between intra- and inter-valley processes~\cite{Schonhoff2016}. Recognizing the importance of the momentum dependence of interactions, here we treat the Coulomb repulsion on the same footing as phonon-mediated attraction, defining a momentum-dependent interaction and summing it on the Fermi surface.

\subsection{Phonon-mediated attraction}
\label{subsec:phonon}

\begin{figure*}
    \centering
    \includegraphics[width=0.9\textwidth]{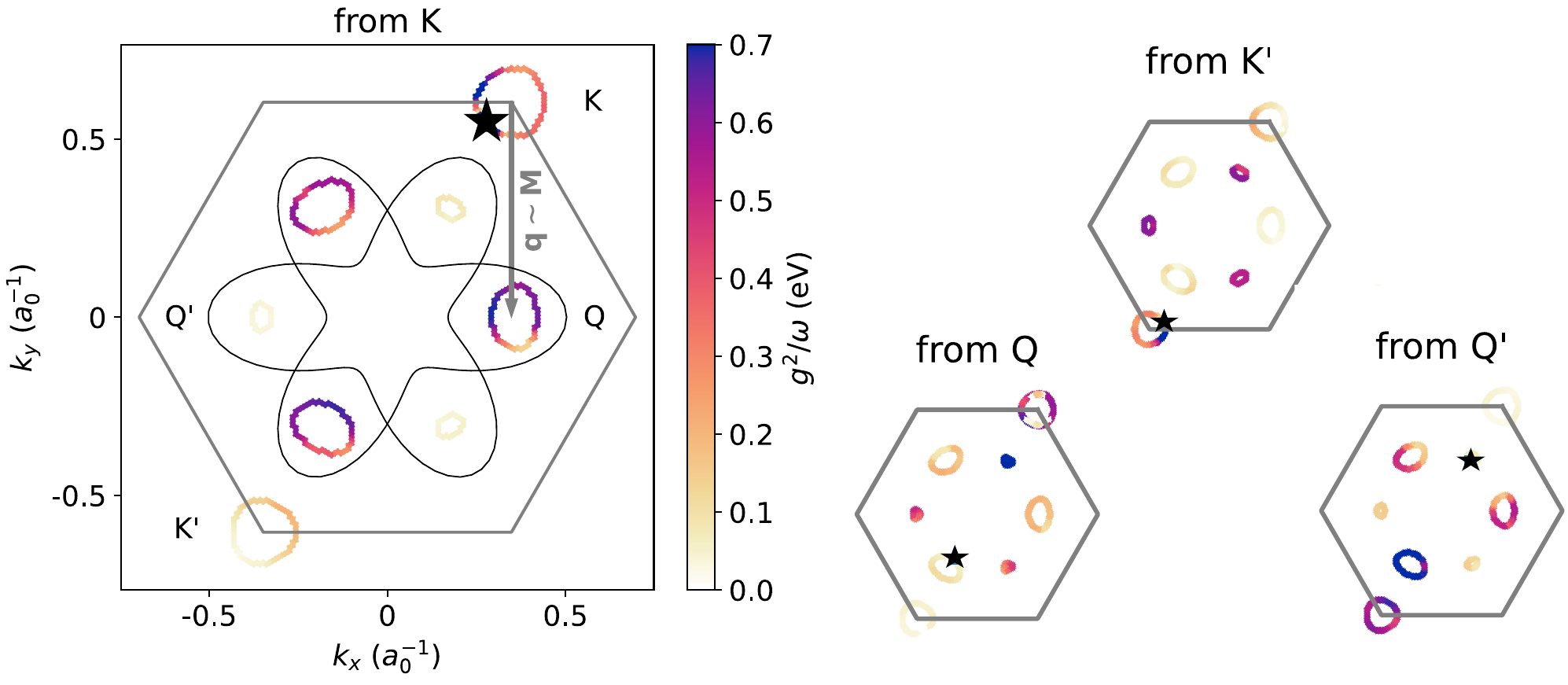}
    \caption{Electron-phonon interactions between states at the Fermi level for MoS$_2$ at a high electron doping of $n=2 \ 10^{14}$ cm$^{-2}$. $a_0$ is the Bohr radius.  
    An initial state $\mathbf{k}_i$ (black star) is picked in one of the 4 valleys and the strength of the phonon-mediated interaction $\sum_{\nu} g_{\mathbf{k}_i,\mathbf{k}_f}^2 / \omega_{\mathbf{k}_f-\mathbf{k}_i, \nu}$ is computed for final states $\mathbf{k}_f = \mathbf{k}_i+\mathbf{q}$ at the Fermi level.  Here the Fermi level crosses all 4 valleys, but at lower doping that might not be the case.}
    \label{fig:MoS2-gooms-e}
\end{figure*}

The quantities needed to estimate the phonon-mediated attraction $G_{vw}$ from first-principles are the values of the squared electron-phonon coupling divided by the phonon frequency, $g^2_{\mathbf{k}\mathbf{k'}, \nu}/\omega_{\mathbf{q}, \nu}$, at the Fermi level and between states in valleys $v$ and $w$.
As we have recently shown \cite{Sohier2019,Sohier2023}, and will confirm in this section, the important feature of electron-phonon interactions in TMDs is the activation of strong intervalley KQ and intravalley KK couplings when carrier density is increased enough to occupy the Q states. 

Here we consider the case of high doping, when the Fermi level crosses all valleys such that all scattering transitions are present.
The values of $g^2_{\mathbf{k}\mathbf{k'}, \nu}/\omega_{\mathbf{q}, \nu}$ between states at the Fermi level are shown in Fig. \ref{fig:MoS2-gooms-e} for MoS$_2$, although similar results are obtained for all other TMDs. Differently from what is typically done in the literature, this is computed by including doping explicitly. This is achieved by performing density functional theory simulations (DFPT) in a field-effect set up that allows for doping~\cite{Sohier2017} and constructing a model that takes into account the unusual dependency of screening on the valley occupations \cite{Sohier2019,Sohier2023}, as detailed in Methods.

To identify the most relevant couplings, Fig.~\ref{fig:MoS2-gooms-e} shows $\sum_{\nu} g_{\mathbf{k}_i,\mathbf{k}_f}^2 / \omega_{\mathbf{k}_f-\mathbf{k}_i, \nu}$  for initial states $\mathbf{k}_i$ in the different valleys as a function of the possible final states $\mathbf{k}_f$ on the Fermi surface. 
For an initial state in K (main panel), intravalley KK and intervalley KQ couplings are stronger than KK' or KQ' by nearly an order of magnitude. 
Moreover, the couplings are rather uniform when considering final states within a given valley, which justifies the approximation of dealing with valley-averaged quantities on the Fermi surface. 

The other smaller panels in Fig.~\ref{fig:MoS2-gooms-e} show the same quantity but for initial states in different valleys. 
This highlights the symmetry in the pairs of strongly coupled valleys.
While the main panel showed that states in K mostly couple to the K and Q valleys, we can now see that states in K' mostly couple to the K' and Q' valleys. Inversely, states in Q (Q') are mainly coupled to K (K').
We can now rationalize the prime notation as a pairing of strongly phonon-coupled valleys (Q with K and Q' with K'). 
QQ' (and Q'Q) couplings are also strong in the case of MoS$_2$. 
However,  the carrier density necessary to reach the Q' valley is not likely to be achieved experimentally. The most relevant result is thus the predominance of intravalley (KK and K'K') and intervalley (KQ and K'Q') couplings with respect to intervalley KK'.

The origin of the strong intravalley KK and K'K' couplings is the A$_1$ optical phonon mode, corresponding to a breathing mode involving the out-of-plane movement of the chalcogens. As shown in \cite{Sohier2019,Sohier2023}, the strong electron-phonon coupling of this mode is not screened by free carriers in the conduction band when the Fermi level is above the Q valley.

The origin of the strong KQ (and K'Q') coupling is a particularly large interaction with the zone border longitudinal acoustic phonon around the $\mathbf{q}=\mathbf{M}$ point. This applies to all the TMDs studied here. This mode is characterized by in-plane displacements of the transition metal combined with out-of-plane displacements of the chalcogens (see, e.g. phonon visualizer in \cite{Campi2022}). This seems to maximize the variations in the angle of the bonds between transition metal (Mo, W) and chalcogens (S, Se). The K and Q valleys correspond to out-of-plane and in-plane orbitals~\cite{Cappelluti2013}, respectively. Strong variations of the bond angle are consistent with strong coupling between in- and out-of-plane orbitals, and transitions between K and Q valleys.

This strong coupling comes with a doping-dependent softening of the LA phonon at $\mathbf{q} = \mathbf{M}$ that is not included in the above results. More precisely, the softening starts when Q is occupied, then gradually increases with doping~\cite{Marini2023}. The softening is difficult to evaluate accurately for different temperatures and dopings, especially due to anharmonic effects~\cite{Marini2023}.
Here, this softening will be approximately included by applying a uniform $30\%$ reduction to the corresponding phonon frequency. This corresponds to the result of the anharmonic calculations in Ref. \cite{Marini2023} for MoS$_2$ at $n \sim 10^{14}$  cm$^{-2}$ where superconductivity is typically found. This only impacts our results when Q is occupied, and we note that the softening could very well be more pronounced at higher dopings.

Fig.~\ref{fig:MoS2-gooms-e} shows electron-phonon couplings at a high electron doping of $n=2 \times 10^{14}$ cm$^{-2}$. For completeness, it is useful to describe the low doping case, when the Q and Q' valleys are not yet crossed by the Fermi level. In that situation, the only relevant couplings involve the K and K' valleys. Comparing with their respective high doping values, the KK' coupling is similar whereas the KK coupling is significantly weaker. In fact, KK and KK' couplings are comparable at low doping. A more quantitative study of doping dependence is performed in Sec. \ref{sec:Mdopdep}.

The data showed in Fig.~\ref{fig:MoS2-gooms-e} can be collected to reconstruct the matrix elements $G_{vw}$ upon averaging over the initial states in valley $v$ and summing over final states in valley $w$, Eq.~\eqref{eq:matrix}. In practice, as already mentioned, the limited variation with initial state within a valley (see also Fig.~\ref{fig:MoS2-GVvski}) makes the averaging procedure not  necessary, in favor of taking just one representative initial state. On the other hand, the sum over final states at the Fermi level is indeed performed. We use a triangular integration technique~\cite{Ashraff1987} rather than replacing the Dirac deltas $\delta(\varepsilon'- \varepsilon_{F})$ by some finite-width distribution which would imply the need for convergence studies.

\subsection{Coulomb repulsion}
\label{subsec:coulomb}

The main ingredient to obtain the Coulomb repulsion contribution $C_{vw}$ from first principles is the screened Coulomb interaction $V(q)$ between states  at the Fermi level in valleys $v$ and $w$. We compute this quantity as a function of the momentum transfer between an initial and a final state, 
$q = |\mathbf{k}_f - \mathbf{k}_i|$, as in Eq.~\eqref{eq:scrCoul}, that is as the bare Coulomb potential divided by the macroscopic static dielectric function $\epsilon(q)$ of the doped TMD. The latter is obtained within the random-phase approximation (RPA), assuming the independent particle susceptibility to be the sum of contributions from the neutral material and from the extra free carriers~\cite{Sohier2023}. 
Concerning the latter contribution, we are mostly interested in the metallic screening effects when $q$ is small enough to correspond to intravalley transitions. 
At large momenta, we assume the dielectric function to be dominated by  interband transitions, i.e. those that make up the dielectric function of the neutral material. Those assumptions are reflected in the choice of a simple complementary error function to model the wavefunction overlaps in the calculation of the free carriers contribution, see App. \ref{app:screened_coulomb}.
Furthermore, we assume an isotropic dielectric function.
Since the Fermi surfaces of all valleys are roughly isotropic, this is a good approximation for small momenta corresponding to intravalley transitions. It is also a good approximation at larger momenta where the interband contributions are assumed to be dominant.

\begin{figure}[h]
    \centering
    \includegraphics[width=\linewidth]{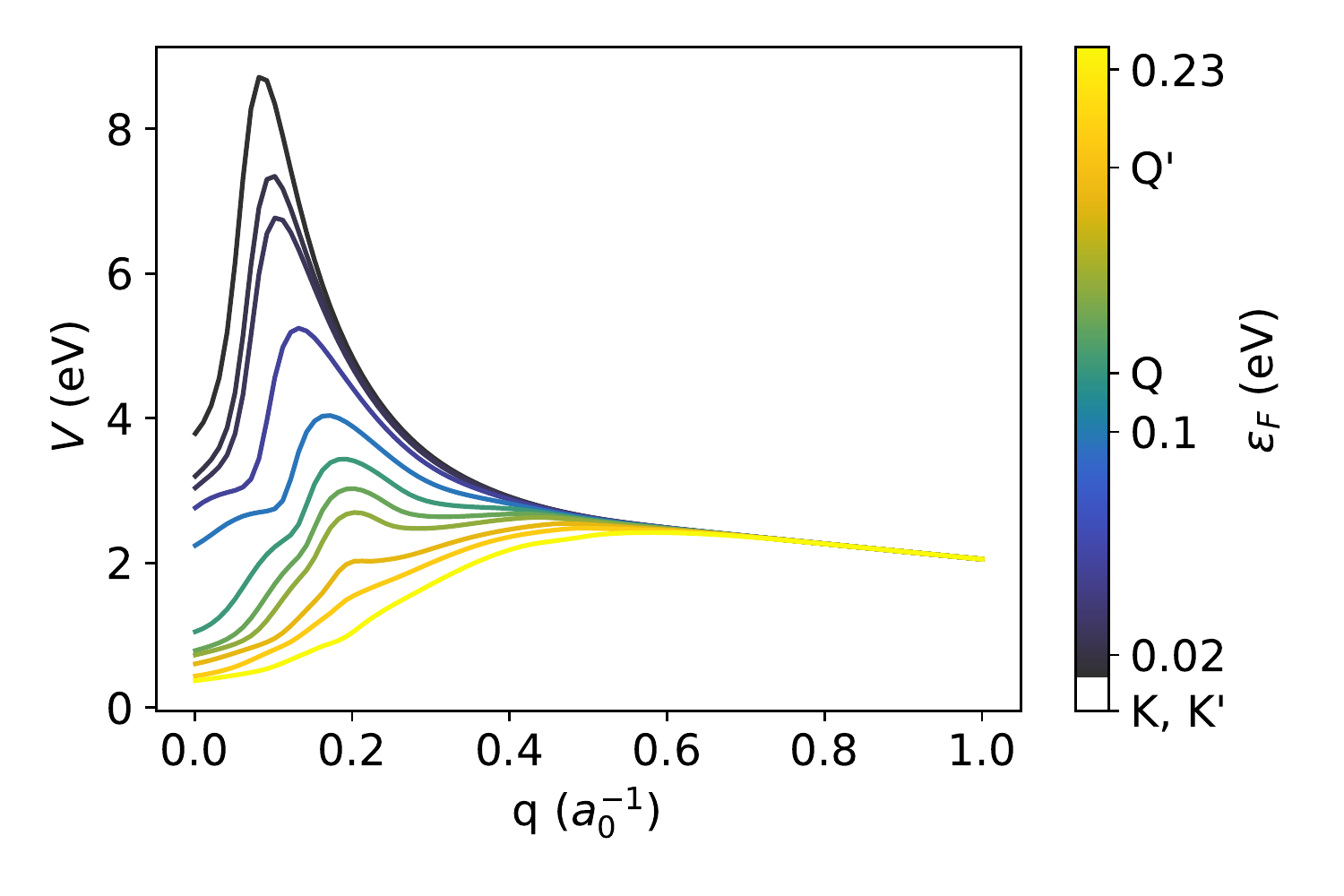}
    \caption{Momentum and doping dependency of the static, screened Coulomb interaction in electron-doped MoS$_2$. $a_0$ is the Bohr radius. The color bar shows the Fermi level, with the bottom of the K' valleys as a reference. The position of the valleys are indicated.}
    \label{fig:Coulomb}
\end{figure}

Fig.~\ref{fig:Coulomb} shows results for the static screened Coulomb interaction  $V(q)$ for different Fermi levels, i.e.\ different doping concentrations. 
Intra- and intervalley processes are characterized by a very different momentum dependence. At small $q$, relevant for intravalley scattering, and in particular for $q<2k_F$ --with $k_F$ being the Fermi wave vector-- there is a strong momentum dependence that arises from the metallic screening associated with free carriers. On the contrary, in the regime of large momentum typical of intervalley processes, and in particular for $q>2k_F$, the screened interaction is slowly varying with $q$, with a behaviour that is reminiscent of 3D dielectric screening, as indeed at large momentum $1/q$ is smaller than the thickness of the 2D layer. 

A distinct behaviour between intra- and inter-valley processes is found also for what concerns the doping dependence. Intravalley coupling is very sensitive to doping as in the long wavelength limit the screened Coulomb potential $V(q \to 0)$ is inversely proportional to the density of states (DOS), as captured already by the Thomas-Fermi approximation. At low doping, only the K/K' valleys are occupied, with a relatively small DOS. On the contrary, at high doping also the Q valleys get occupied, with a much larger DOS (both because of the higher degeneracy and an heavier effective mass), with a significant suppression of the screened interaction. The suppression associated with metallic screening extends to finite $q$ but still within the Fermi surface, with a flattening of $V(q)$ and a reduction of Coulomb repulsion. 
Intervalley interactions characterized by large momentum transfer $q$ are not affected by metallic screening, and the Coulomb repulsion is doping-independent. 

The screened Coulomb interaction is then summed over possible final states on the Fermi surface for a given valley to obtain the Coulomb matrix elements $C_{vw}$ through Eq.~\eqref{eq:matrix}.

\section{Results}
\label{sec:results}

The procedure outlined above to compute $G_{vw}$ and $C_{vw}$
is repeated for different doping concentrations and in different materials. The linearized form of Eq.~\eqref{eq:eigen} is then solved to find the corresponding $T_c$, and from the eigenstates we gain information on the nature of the superconducting state. 
However, before illustrating the main results, it is important to stress that the gap equations depend on the value of the $\alpha$ parameter, which tunes the relative strength of Coulomb repulsion and phonon-mediated attraction. 
The model in Sect.~\ref{sec:model} is intended to describe phonon-driven superconductivity, with Coulomb repulsion as a perturbation. 
The starting point is thus $\alpha = 0$, corresponding to the complete neglect of Coulomb interactions.
In that case the gap equations are fully determined by phonon attraction and any superconducting solution found in this regime is obviously purely phonon-driven.
This is still true for finite but small values of $\alpha$. 
In the opposite limit of large $\alpha$, all $C_{vw}$ matrix elements eventually become larger than the corresponding $G_{vw}$, meaning that the valley couplings are all repulsive, and the gap equations, as written in Eq.~\eqref{eq:eigen} assuming  a dominating phonon attraction,  fail to be  valid. 

We thus restrict our investigations to a maximum value of $\alpha=0.3$. Although there are small variations among TMDs, this roughly corresponds to the point where the strongest phonon-mediated coupling ($G_{KQ}$) is half compensated by the Coulomb repulsion ($M_{KQ}=G_{KQ}-\alpha C_{KQ} \sim G_{KQ}/2$), while the second strongest ($G_{KK}$) is fully compensated ($M_{KK} \sim 0$), all other elements of  $M$ being already negative. Note that those criteria only make sense when Q is occupied, but as we'll see, that is a necessary condition for superconductivity anyway, even for $\alpha=0$. 
The Coulomb repulsion was simply computed from the screened Coulomb interaction for single-layer TMDs in vacuum. One may consider its reduction by $0\leq \alpha \leq 0.3$ as including the contribution from a dielectric environment. In particular, in the common case of an ionic gate, the liquid's highly polarizable ligands are very close to the layer, closer than the average distance between the added carriers in the conduction band. 

In the following we set $\alpha = 0.2$ when studying the doping dependence of the interaction matrix elements. The observed trends are actually robust for a wide range of $\alpha>0.05$, and 
$\alpha = 0.2$ is simply chosen to make the effects of Coulomb repulsion quantitatively obvious, without reaching the maximum considered as the threshold for validity ($\alpha = 0.3$). 
The solutions to the gap equations are then investigated for different doping levels and at 5 representative values in the entire range of $\alpha$ ($= 0, 0.05, 0.1, 0.2, 0.3$). 

\subsection{Doping dependence of interaction matrix elements}
\label{sec:Mdopdep}

\begin{figure}[h!]
    \centering
\includegraphics[width=\linewidth]{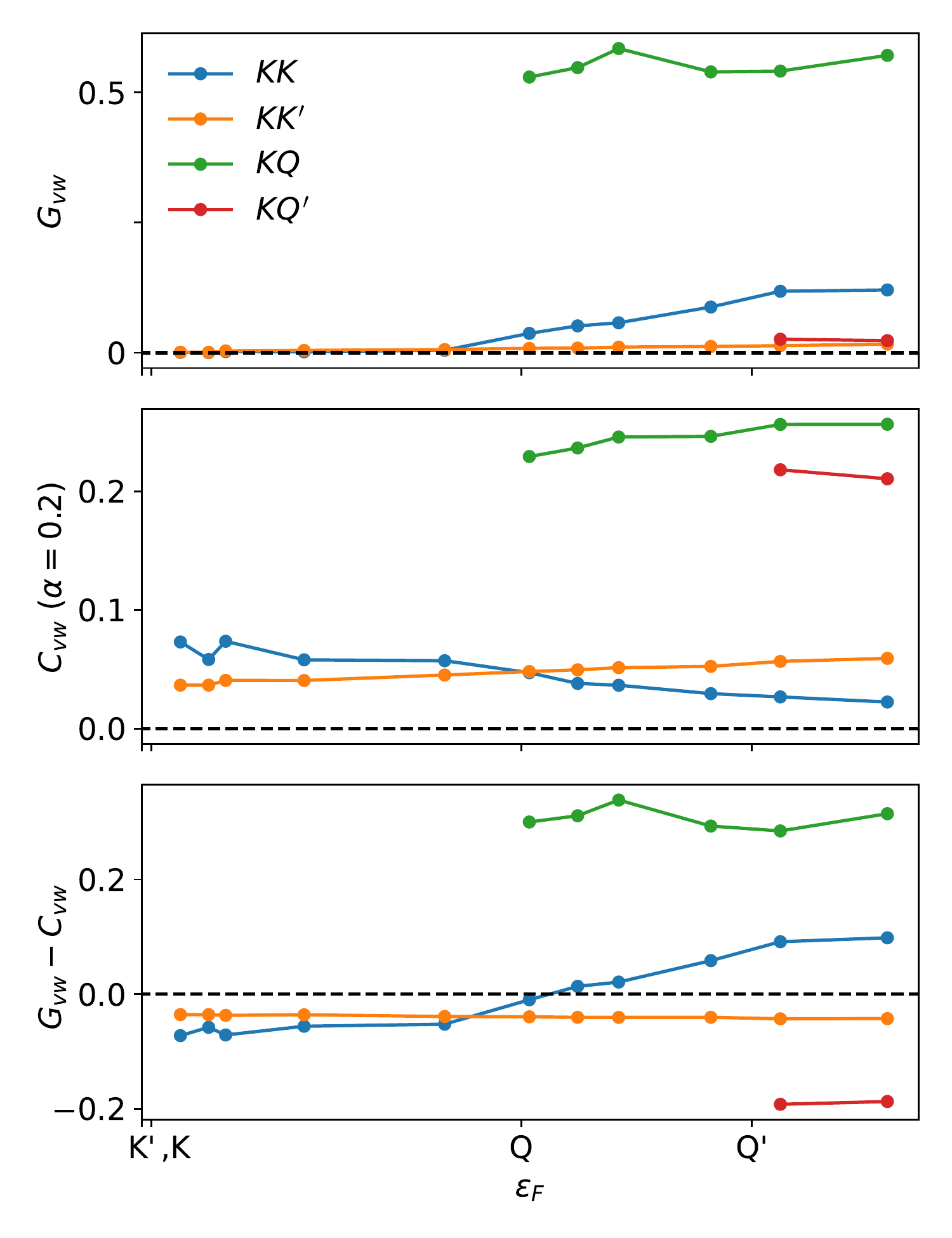}
    \caption{Evolution of the intervalley phonon-mediated attraction, Coulomb repulsion, and their difference with electron-doping in MoS$_2$. 
    The  horizontal axis shows when the Fermi level crosses the bottom of each valley. The values of the Fermi levels and carrier densities corresponding to each point can be found in Fig.~\ref{fig:Tc_Ef}. Here, the tuning prefactor $\alpha$ for the Coulomb repulsion is set at $0.2$. 
    Only active couplings--i.e.\ between valleys that are populated at that Fermi energy--are reported.
    }
    \label{fig:MoS2-GC}
\end{figure}

The matrix elements of Eq.~\eqref{eq:matrix} describe phonon-mediated attraction and Coulomb repulsion. Fig.~\ref{fig:MoS2-GC} reports representative results for the case of MoS$_2$, for $\alpha = 0.2$, showing the relative importance of those two contributions as a function of doping. For simplicity, we plot only the matrix elements with K as a first index. Those, along with the symmetry of the electron-phonon  and Coulomb interactions ($\frac{g^2_{kk+q}}{\omega_{q, \nu}}$ and $V(q)$) with respect to valley indices, are enough to describe the trends for the full matrix.

The top panel of Fig.~\ref{fig:MoS2-GC} shows the evolution of $G_{K,K}$, $G_{K,K'}$, $G_{K,Q}$, and $G_{K,Q'}$ as a function of the Fermi energy. 
At low doping, the Fermi level does not cross the Q valley, such that only matrix elements involving K and K' can be non-zero. Those are finite but very small ($<0.01$) on the scale of  Fig.~\ref{fig:MoS2-GC}.

Interestingly, however, intravalley KK (and K'K') coupling becomes strong  as Q is occupied, with an enhancement by more than an order of magnitude. This is the consequence of the peculiar multivalley screening effect discussed earlier by some of us~\cite{Sohier2019,Sohier2023}. 
In particular, electron-phonon coupling is dominated by the A$_1$ phonon mode. The corresponding coupling mechanism can be thought of as a shift in energy of the K and Q valleys following the periodic phonon displacement pattern and out-of-phase with one another. 
The relative amplitude of the K and Q shifts are such that the charge density perturbations associated to each valley, opposite in sign, are equal in amplitude. Thus, they compensate each other. With a net charge density perturbation that vanishes, the screening from the conduction band electrons also vanishes when Q is occupied, leading to a large electron-phonon coupling.

Intervalley KQ scattering is obviously present only when doping is high enough for the Fermi level to cross the Q valley. Once active, it is an order of magnitude stronger than the others. 
We have verified numerically that, as expected from symmetry, primed valleys K' and Q' also couple strongly to each other. 
The appearance of the large phonon-mediated couplings discussed earlier, which follow the prime notation, thus coincides with the occupation of the Q valley.

The middle panel of Fig.~\ref{fig:MoS2-GC} reports analogous results for Coulomb coupling matrix elements $C_{K,K}$, $C_{K,K'}$, $C_{K,Q}$, and $C_{K,Q'}$, with the aforementioned choice of $\alpha = 0.2$.
The intervalley (KK') component is only slightly dependent of the Fermi level location. This is consistent with the constant $V(q)$ in Fig.~\ref{fig:Coulomb} associated to a slight increase of the density of states as the Fermi level increases. The intravalley component $C_{KK}$ instead gradually decreases with doping, from a factor of two above the intervalley $C_{KK'}$ to slightly below it.  When the doping level is increased so that also the Q valley is occupied, the suppression of $C_{KK}$ is slightly accelerated.
More importantly,  intervalley couplings to the Q valley become active, and they are significantly larger than the other matrix elements. This is because the main ingredient determining the relative magnitude of the matrix elements from valley to valley is the DOS in the final valley $w$. In particular, the larger DOS of Q and Q' makes $C_{v,Q}$ and $C_{v,Q'}$ matrix elements significantly larger. Importantly, the Coulomb matrix elements are not sensitive to the primed/unprimed distinction between the valleys. Not only KK and KK' matrix elements have comparable orders of magnitude, but also KQ and KQ' components have similar values. 

Plotting $G_{vw}-C_{vw}$ in the bottom panel of Fig.~\ref{fig:MoS2-GC} highlights some interesting features of the competition between phonon-mediated attraction and Coulomb repulsion. Indeed, some elements of $M$ change sign as a function of doping.
At small carrier concentration, the matrix elements are either null when the corresponding intervalley transition is not on the  Fermi surface, or small and negative when Coulomb repulsion dominates. As Q is occupied, both $G_{KQ}$ and $G_{KK}$ increase strongly, overcoming the Coulomb repulsion and leading to positive $M_{KQ}$ and $M_{KK}$ elements. As we will see in the next section, superconductivity emerges around those conditions. 

Phonon-mediated attraction does not overcome Coulomb repulsion for all valley pairs: $M_{KK'}$ and $M_{KQ'}$ remain negative. Strong phonon-mediated couplings follow the prime notation, while the Coulomb repulsion is similar among and between primed and unprimed valleys, and varies mostly according to the density of states associated with the second valley index.
As a result, we observe different outcomes of their interplay in different valley pairs.
In particular, once Q is occupied, phonon mediated attraction systematically prevails among primed or unprimed valleys, while the KK' coupling remains repulsive. The KQ' coupling is also negative when active, although it is less relevant in practice since the Q' valley is unlikely to be occupied in typical experiments. 
The main effect of introducing Coulomb repulsion is thus to invert the sign of the valley couplings where phonon-mediated attraction is weak, while keeping the couplings attractive between the most strongly phonon-coupled valleys.

The pivotal sign inversion of the KK' coupling, illustrated here with $\alpha=0.2$, actually happens over a broad range of values of $\alpha$, starting already from $\alpha\sim 0.05$ (depending on the TMD and doping level). This is ultimately due to the KK' phonon-mediated attraction being quite weak for any doping level. 
For values of $\alpha$ above the chosen threshold of $0.3$, the dominance of phonon mediated attraction is challenged even for KK and KQ couplings, a regime which is beyond the scope of this work. 
As we will see in the next section, the dominance of KK and KQ phonon-mediated attraction as a function of doping, i.e. when Q is occupied, and the possible coexistence of attractive and repulsive couplings between different valley pairs are the key features to understand the onset and nature of superconductivity in TMDs. 

\subsection{Superconducting solutions}

\begin{figure*}
    \centering
    \includegraphics[width=0.98\textwidth]{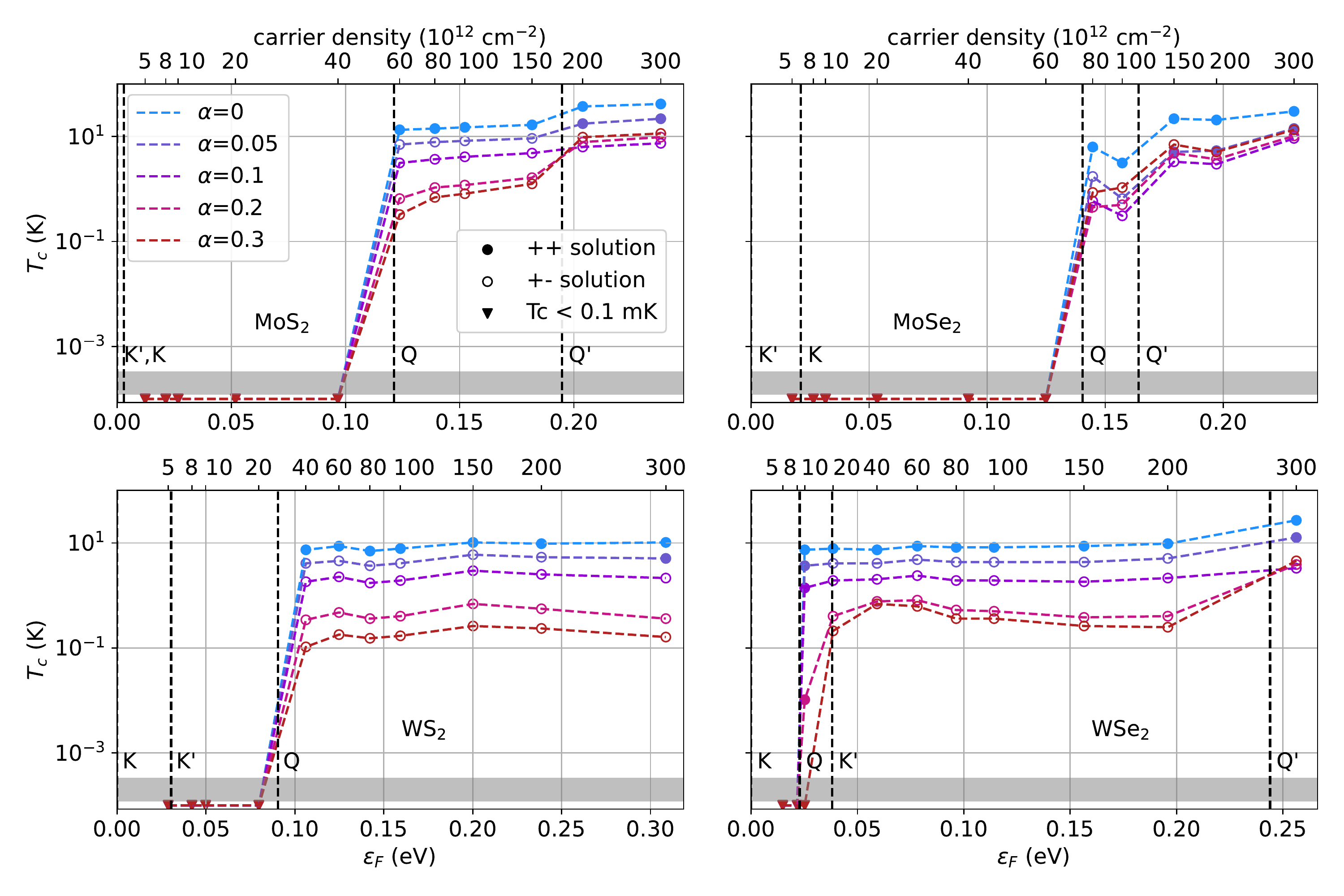}
    \caption{Critical temperature as a function of Fermi level (with respect to the bottom of K valley) for the four TMDs, for different values of the prefactor to Coulomb repulsion, $\alpha$ . The top axis indicates the electrostatic doping level corresponding to each point. Markers indicate the type of solution found. The Q and Q' valleys of MoS2 have been lowered by $100$ meV with respect to the DFT results.}
    \label{fig:Tc_Ef}
\end{figure*}

With the $G$ and $C$ matrices computed, the linearized form of Eq.~\eqref{eq:eigen} is solved for all four TMDs considered here (MoS$_2$, MoSe$_2$, WS$_2$, and WSe$_2$), at different carrier concentrations, and for four representative values of $\alpha$ (0.0, 0.05, 0.1, 0.2, 0.3) within the range discussed above.
The critical temperature $T_c$ is  found as the lowest temperature for which Eq.~\eqref{eq:eigen} allows for a non-trivial solution. We only look for solutions with a finite and experimentally relevant $T_c>10^{-4}$ K. 
The behavior of $T_c$ as a function of doping and the nature of the solution is determined by the interplay between attractive phonon-mediated and repulsive Coulomb interactions. This interplay is different for each pair of valleys within the multivalley Fermi surface, and strongly depends on doping.
A summary of all corresponding results is reported in Fig.~\ref{fig:Tc_Ef}.

\subsubsection{Onset of superconductivity}

There are several robust features common to all TMDs studied here.  The onset of a critical temperature above the chosen threshold for experimental significance ($>0.1$ mK) systematically coincides with the Fermi level crossing the bottom of the Q valley. This is true whatever the value of the tuning parameter for the Coulomb repulsion, $\alpha$. 
This  holds in spite of the variations in the carrier density at which Q is reached, i.e.\ it does not depend on the energy of the bottom of the Q valley with respect to the K and K' valleys.
Clearly the emergence of phonon-driven superconductivity is triggered by the activation of KQ and the enhancement of KK couplings as the Q valley is occupied.

Once the Q valley is occupied, we find a fairly constant $T_c$ of a few Kelvins for all TMDs. 
This value is maximal in the absence of Coulomb repulsion ($\alpha=0$) and gradually decreases as $\alpha$ is turned on, in a similar way for all TMDs. For the maximal value of $\alpha=0.3$, $T_c$ is decreased by an order of magnitude. 

The precise value of $T_c$ and its doping-dependence are quite sensitive to the softening of the LA phonons at $\mathbf{q} = \mathbf{M}$, which is delicate to evaluate ab initio, notably due to anharmonic effects~\cite{Marini2023}. Indeed, the associated phonon-mediated coupling between the K and Q valleys is very large and drives the superconductivity.
As already mentioned, for this work we simply apply a $30\%$ reduction of the LA(M) frequency, based on the results of Ref.~\cite{Marini2023}. Realistically, one expects a doping-dependent softening, starting from the occupation of the Q valley and steadily increasing, likely past the $30\%$ used here, with a corresponding increase in $T_c$. Nevertheless, we expect the current treatment to give a good  estimation in the relevant doping range $n = 10^{13} - 10^{14}$ cm$^{-2}$. Most importantly, it is sufficient to understand the underlying physics.

Thus, despite variations in their chemical composition, electronic bands and phonon properties, all the semiconducting TMDs considered here consistently display phonon-driven superconductivity with a $T_c$ on the order of a few Kelvins, the onset of which systematically coincides with the occupation of the Q valley.

\subsubsection{Emergence of exotic solutions}

While superconductivity is driven by electron-phonon interactions, the Coulomb repulsion plays an important role. Interestingly, even at small tuning parameter $\alpha$, it modifies the coupling between valleys significantly enough to change the nature of the superconducting solutions, as characterized by the eigenvector $\Delta_v$ corresponding to a unit eigenvalue in the linearized gap equation~\eqref{eq:eigen}. Its components provide information on the relative sign and magnitude of the superconducting order parameter in each  valley. In Fig.~\ref{fig:Tc_Ef}, this is illustrated by a switch from filled to empty markers, used to denote qualitatively different solutions. This consistently happens for all finite values of $\alpha$ considered.

\begin{figure}
    \centering
    \includegraphics[width=0.48\textwidth]{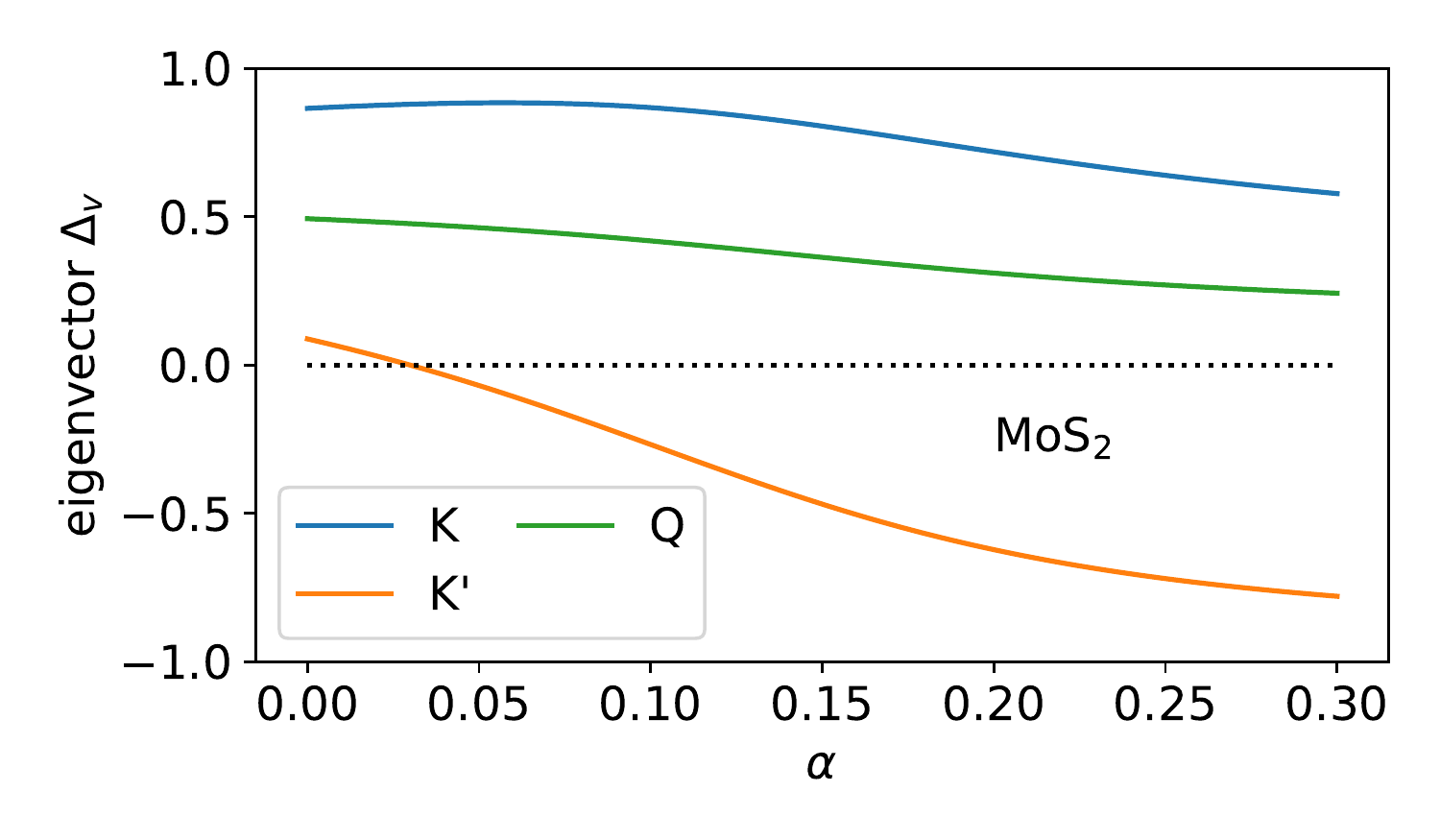}
    \caption{Components of the eigenvector $\Delta_{v}$ corresponding to the superconducting solution as a function of the Coulomb prefactor $\alpha$, in MoS$_2$. The electron doping is fixed and corresponds to the second data point above Q, with $\varepsilon_F \approx 0.14$ eV, and $n=8 \times 10^{13}$ cm$^{-2}$. }
    \label{fig:eigenvec}
\end{figure}

To investigate further the change in the nature of the solution, Fig.~\ref{fig:eigenvec} shows the components $\Delta_v$ as a function of $\alpha$. Doping is chosen to correspond to the standard experimental conditions in which superconductivity is observed, when the Fermi level is about $\sim 10$ meV above Q.
For $\alpha = 0$, all components of $\Delta_v$ are positive. The K valley dominates, followed by Q, while K' has a small but positive contribution. 
For fairly small values of $\alpha \lesssim 0.05$, the K' component goes through zero then changes sign. This coincides with the change of sign for the $G_{KK'}-C_{KK'}$ matrix elements mentioned before.
The K and Q components stay positive. 
Conventional phonon-driven superconductivity corresponds to a positive order parameter in all valleys, as obtained for $\alpha=0$. The introduction of Coulomb repulsion thus systematically leads to the appearance of exotic solutions, i.e. an eigenvector with components of different signs, which corresponds to the order parameter changing sign for electrons in different valleys.
Note, however, that the solutions we find do not break the point-group symmetry ($s$-wave), which in conventional superconductors, such as lead or aluminum, ensures robustness with respect to disorder \cite{anderson1959theory}.

A necessary (but not sufficient) condition for the existence of those exotic solutions is to have coexisting attractive and repulsive elements in the matrix $M$. The strength of Coulomb relative to the phonon-mediated interaction determines the sign of the coupling between each pair of valleys. Depending on momentum, doping, and  $\alpha$, the repulsive Coulomb term $C_{vw}$ overcomes the phonon-mediated term $G_{vw}$ in some but not all of the matrix elements.  Due to their strong phonon-mediated coupling, K and Q valleys remain attractively coupled. The opposite-sign channels are made up of (KQ) and (K'Q'). Those channels are attractive within themselves, but repulsive between each other. 
This corresponds to a pairing parameter that has opposite signs in the K and K' valleys. The Q (Q') valley shares the sign of the K (K') valley, although with a smaller magnitude.

Going back to Fig.~\ref{fig:Tc_Ef}, note that in the peculiar case of WSe$_2$ where the Q valley is reached before the K' valley, the standard solutions persists for all values of $\alpha$ when Q is occupied but K' is not. This is consistent with K' (and Q' when applicable) being the negative channel. Exotic solutions only emerge once K, K' and Q are occupied.

\subsubsection{Summary of the results}

The main findings of this work with respect to the current understanding of superconductivity are twofold. First, while the onset of superconductivity at Q valley occupation was already anticipated~\cite{Piatti2018,Sohier2019,Piatti2019,Ding2022}, the underlying physical mechanism is unambiguously clarified and shown to be robust.
Second, we propose a novel paradigm for the nature of superconductivity, with a phonon-driven mechanism leading to exotic solutions in the presence of Coulomb repulsion.

The emergence of exotic solutions is an important result because their properties differ from the standard solutions, leading to different implications for experimental observations. 
The sign change in the order parameter while going between different valleys allows the system to take advantage of the intervalley Coulomb repulsion, which can dominate the phonon-mediated attraction for large enough $\alpha$. The overall superconducting state is still fully gapped; however, such a state with sign-changing order parameter is more fragile against atomic-scale disorder. That is because  disorder can scatter electrons between regions with different gap signs, thereby reducing $T_c$ and introducing a finite density of states near the Fermi energy \cite{mazin2008unconventional, fernandes2022iron}.

\section{Comparison/Implications for experiments}
\label{sec:final}

The results of the theoretical analysis presented above are in overall agreement with the main  experimental results reported on different semiconducting TMDs, and provide clear indications to explain aspects of the experiments that are currently not understood. In this Section we discuss experimental observations in the context of the framework provided by the theoretical model discussed in the paper.

A first essential point of our model, is that it predicts that robust superconductivity occurs when the Q-valleys start to be populated with electrons, and that electron-phonon interaction is the main driving force for the superconducting transition. Both the role of Q-valley filling and of electron-phonon interaction have been established by experiments over the last few years. Piatti \etal~\cite{Piatti2018} first and D.D.\ Ding \etal~\cite{Ding2022} later have performed gate-dependent transport measurements (respectively on MoS$_2$ multilayers and WS$_2$ monolayers), showing that  the onset for superconductivity indeed corresponds to filling of the Q-valley. In independent Raman spectroscopy measurements  on monolayers of different semiconducting TMDs performed as a function of electron density~\cite{Sohier2019},  it was further shown  that populating the Q-valley with electrons leads to a drastic increase in the strength of the electron-phonon coupling. This drastic increase --reproduced by our simulations and discussed here in Section~\ref{sec:Mdopdep}-- is indeed the mechanism driving superconductivity. We conclude that  both the assumptions underlying our model, and the result of our calculations are fully consistent with experiments, and provide a microscopic explanation for the key aspects of the gate-induced superconducting transition that are common to the different TMDs investigated. Indeed, not only our calculations show that the superconducting transition is brought into an experimentally accessible range of critical temperatures when the Q-valley is filled, but also they give a range of critical temperatures for the different TMDs that matches the order of magnitude observed experimentally. 

The second important conclusion of our model is that superconductivity --despite being driven by electron-phonon interaction-- has a $s_{+-}$ character --i.e., an order parameter with different sign in the different populated bands/valleys-- due to the non-negligible role of electron-electron interaction. We argue that the $s_{+-}$ nature of  the superconducting state is essential to interpret different aspects of the experimental results that are currently not understood. That is because in a $s_{+-}$ state disorder has very different physical effects as compared to conventional superconducting states (even multiband ones), in which the order parameter has the same sign everywhere on the Fermi surface. The situation has been summarized in detail by Mazin and Schmalian~\cite{Mazin2009}, in their review of pnictide superconductors (many of which are also considered to host a  $s_{+-}$ superconducting state) that we follow here to guide our discussion.

How disorder affects $s_{+-}$  superconductors depends strongly on the nature and the density of defects present in the system. Strong and short-range disorder potentials (such as those caused by atomic vacancies) introduce sub-gap states, without strongly affecting the superconducting transition temperature. This is theoretically predicted to happen up to rather large densities of defects, sufficient to suppress  the "hard" gap of the pure material by filling it with states  at all energies (all the way to $E=0$, i.e. at the Fermi energy). The situation is different for weak, long-range disorder potentials, such as those caused by inhomogeneity in a background charge distribution (e.g., as the one generated by spatial fluctuations in the density of dopants, in a common doped semiconductor). In   $s_{+-}$ superconductors, such weak and long-range disorder potentials suppress the critical temperature and eventually kill superconductivity. 

For gate-induced superconductivity in TMDs, both types of disorder are normally present. 
Chalcogen vacancies, which are unavoidably present in semiconducting TMDs with densities that depend on the specific compound, naturally lead to strong, short-range potentials. In MoS$_2$, for instance, it is common to have up to $10^{13}$ cm$^{-2}$ Sulfur vacancies;  in other  TMDs the density of these vacancies is likely smaller, but still sizable. Based on the general behavior of $s_{+-}$  superconductors discussed above, these chalcogen vacancies create sub-gap states and can explain why a large density of states (DOS)  at sub-gap energies has been observed in experiments on  MoS$_2$ , which would not be possible to explain in terms of a usual fully gapped $s$-wave superconducting state. To substantiate more this conclusion, it will be important in the future  to  measure the sub-gap DOS  in other TMDs with lower density of chalcogen vacancies, or to find ways to correlate  the sub-gap DOS  with the density of these vacancies.

Weak long-range disorder, instead, originates from  potential fluctuations associated to variations in the density of ions in the ionic liquid used for electrostatic gating. In a conventional $s$-wave superconductor, these potential fluctuations would have no important physical effect, but  in a $s_{+-}$  they can suppress $T_c$ and weaken the superconducting state. We believe that this may be the reason for the superconducting dome that is seen in some experiments. Indeed, in some case (e.g., MoS$_2$), a dome with consistent properties has been reported in experiments by different groups, on identical kinds of devices. In other experiments, however, whether a dome is present or not seems to depend on the  quality of the devices investigated (as measured, for instance, by the value of the electron mobility).  In particular, in WS$_2$ monolayers, devices realized on hBN  substrates exhibit larger $T_c$ and no dome (data are compatible with a saturation of $T_c$), whereas in devices with the WS$_2$ monolayer  in direct contact with a SiO$_2$ substrate the critical temperature is lower, and a clear dome-shaped dependence on electron density is observed. The latter experiments --as well as a comparison of reported data (see figure in Calandra's paper)-- suggest that the dome-like dependence of $T_c$ on electron density is an extrinsic phenomenon. This is consistent with the predictions of our model, namely that in the absence of disorder $T_c$ approximately saturates upon increasing electron density, once the Q-valley starts to be populated, and that the dome may be understood as an extrinsic effect due to weak, long-range disorder due to the $s_{+-}$ nature of the superconducting state. 

In summary, known experimental facts are consistent with all basic aspects of the behavior predicted by our model and the $s_{+-}$ nature of superconductivity provides a scenario that  allows us to rationalize phenomena that were so far difficult to understand theoretically.

\section{Conclusion}
We develop a model of superconductivity in semiconducting TMDs that is sophisticated enough to capture the momentum dependency of phonon-mediated and Coulomb interactions within a complex multivalley structure, yet simple enough to draw clear physical insight.
We find systematic, robust results that apply to all the TMDs studied here, i.e.\ electron-doped MoS$_2$, WS$_2$, MoSe$_2$, and WSe$_2$.
We find that  superconductivity is phonon-driven, with a doping-dependent onset corresponding to strong electron-phonon interactions being activated as the Q valley is occupied. In particular, zone border acoustic phonons mediate a strong coupling between K and Q valleys, while intravalley couplings are enhanced by a vanishing screening of the zone center A$_1$ mode.
While phonons drive the emergence of superconductivity, Coulomb repulsion plays a crucial role as it is shown to change the nature of the superconducting solution. In particular, the introduction of Coulomb repulsion, even relatively weak, leads to exotic s$_{+-}$ solutions where the order parameter changes sign between valleys.
All robust experimental facts currently reported in the literature are consistent with the results of our model, and the emergence of s$_{+-}$ solutions explains some aspects that were puzzling up to now.
The physical insight gained on the nature and mechanism of superconductivity in TMDs can be used to better control it. For example, one could imagine tuning the Coulomb repulsion via dielectric engineering of the environment to cross over from standard to exotic superconducting solutions.
The model is also general enough to draw conclusions on different systems. First, for hole doping of the same TMDs, we can expect similar qualitative conclusions with the $\Gamma$ valley playing the role of the Q valleys. The fact that the $\Gamma$ valley is much more difficult to reach with hole doping would then explain the lack of experimental evidence for superconductivity on the hole side.
Finally, a similar modelling procedure can be applied to other multivalley 2D materials, leveraging the general strength of intervalley electron-phonon coupling to unveil new superconducting systems.

\begin{acknowledgments}
AFM acknowledges financial support from the Swiss National Science Foundation under project 200020\_178891 and the EU Graphene Flagship project for support. MG acknowledges support from Ministero Italiano dell'Universit\'a e della Ricerca through the PNRR project ECS\_00000033\_ECOSISTER and the PRIN2022 project SECSY. IM acknowledges support by the US Department of Energy, Office of Science, Basic Energy Sciences, Materials Sciences and Engineering Division.    
\end{acknowledgments}

\section{appendix}
\label{sec:methods}

\subsection{Multivalley gap equation}

In this section we elaborate on our treatment of superconductivity. It is a simple generalization of the standard BCS treatment (see for instance \cite{mahan2013many}).
The standard BCS gap equation is:
\begin{equation}
\Delta(k) = \sum_q W(q )\frac{\Delta(k-q)}{E_{k-q} }\tanh{\frac{E_{k-q}}{2k_BT}}.
\end{equation}
Here $E_k = \sqrt{\varepsilon_k^2 + \Delta_k^2}$ is the Bogoliubov quasiparticle energy. The gap function $\Delta(k)$ corresponds to the pairing between spin-up electrons with momentum $k$, and spin-down electrons with momentum $-k$. The interaction $W(q)$ is assumed to be operational only within a window of energies $\pm \omega_D$ around the Fermi level (In BCS, this is because the phonon-mediated interaction is attractive below the phonon characteristic phonon frequency, and decays rapidly above that frequency). 

If the gap function can be assumed to be momentum-independent,  $\Delta$, the equation for the superconducting transition temperature $T_c$ becomes
\begin{equation}
\begin{aligned}
\Delta &= \sum_q W(q )\frac{\Delta}{E_{k-q} }\tanh{\frac{E_{k-q}}{2k_BT_c}}\\
&\approx \Delta \ln\left[ \frac{\omega_D}{k_B T_c} \right]\left\langle \sum_{q} W(q)   \delta(\varepsilon_{k+q} - \varepsilon_F) \right\rangle_{k \in FS}
\end{aligned}
\end{equation}
In our case, there are several symmetry-unrelated Fermi sheets, which in general will have different gap functions. For spin-up electrons, we label them $K,\ K',\ Q,\ Q'$ (see Fig. 2). 
Thanks to the time-reversal symmetry, for spin-down electrons, the pockets have the same shape but are reflected as $k\to -k$. 
Note also that there are three symmetry related $Q$ and $Q'$ Fermi surfaces, but only one $K$ and $K'$. Given that all these Fermi sheet are much smaller than the size of the Brillouin zone, we can safely assume that within each Fermi pocket the gap function can be well approximated by just a single value, $\Delta_v$. 

These gap functions combine into a matrix gap equation that describes their couplings, 

\begin{equation}
\Delta_v \approx \sum_w \Delta_w \ln \left[\frac{\omega_{vw}}{k_B T_c}\right] \left\langle \sum_{q, k+q\in w} W(q)   \delta(\varepsilon_{k+q} - \varepsilon_F) \right\rangle_{k \in v} .
\end{equation}

Since there are three distinct symmetry-related $Q$ and $Q'$ pockets, it is conceivable that the order parameter could have a non-zero angular momentum. However, that would imply that the $K$ and $K'$ pockets remain non-superconducting. This is not energetically favorable; thus, we assumed that all $Q$ (and $Q'$) pockets have the same value of gap
function. Consequently, there are four distinct values of the gap, which leads to the system of four equations, Eq. 1 and Fig 2. Both the interacting strength and the attraction frequency cut off for every pair of valleys are chosen based on the first principles simulations, which we describe next.

\subsection{First-principles simulations}

\subsubsection{Electron-phonon interactions}
The electron-phonon coupling matrix elements are obtained using the method described in Ref. \cite{Sohier2023}, starting from the same ab initio simulations (the results of which were provided there). We report some aspects of this method here for completeness. Some repetition of the content of Ref. \cite{Sohier2023} is inevitable.

The ground state and phonon properties of each semiconducting TMD (MoS$_2$, MoSe$_2$, WS$_2$, WSe$_2$) are simulated within density functional (perturbation) theory using the \textsc{quantum ESPRESSO} package\cite{Giannozzi2017,giannozzi_quantum_2009}. The full simulation is only performed at one carrier concentration of $n=5\times 10^{12}$ cm$^{-2}$, on the relaxed structures. We use norm-conserving, fully-relativistic pseudopotentials with Perdew-Burke-Ernzerhof (PBE) functionals \cite{Hamann2013} from the pseudo-DOJO library \cite{vanSetten2018}, with kinetic energy cutoffs of 50 (WS$_2$), 70 (MoS$_2$) or 80 (WSe$_2$, MoSe$_2$). We use 2D boundary conditions and symmetric gates to induce doping \cite{Sohier2017}.
The electronic momentum grid is non-uniform, with a sampling ranging from $12\times 12$ to $96 \times 96$ around the Fermi surface. The electronic occupations follow the Fermi-Dirac distribution taken at room temperature, with a Fermi level such that the carrier density in the conduction band is $n=5\times 10^{12}$ cm$^{-2}$.
The electron-phonon coupling matrix elements involve initial and final electronic momenta, assumed to be confined within a certain energy window, from the bottom of the conduction band up to around $0.3$ eV above, as represented in Fig. \ref{fig:SpinValleyStructure}. Final states are sampled on a $120 \times 120$ Monkhorst-Pack grid, while the initial states are sampled on a grid twice as coarse. The phonon momenta to compute are obtained by linking initial states to final states. 
The electron-phonon coupling matrix elements at other carrier densities are evaluated using the model of Ref. \cite{Sohier2023}. Notably, this accounts for the variations in free-carrier screening of the intravalley electron-phonon interactions. Indeed, as shown experimentally and explained qualitatively in Ref. \cite{Sohier2019}, the occupation of both K and Q valleys leads to a suppressed screening and thus an enhancement of the intravalley electron-phonon couplings in TMDs. We then modeled and quantified this peculiar screening mechanism as a function of Fermi level and valley occupations in Ref. \cite{Sohier2023}. Standard free-carrier screening is also accounted for.
We effectively end up with a model of electron-phonon interactions as a function of both the Fermi level and the relative position of the Q(Q') valleys with respect to the K valleys. 

Note that the position of the Q valley for MoS$_2$ in the calculations of Ref. \cite{Sohier2023} was fairly high ($> 0.2$ eV). For convenience, and since we do not claim nor aim for accuracy in the relative positions of the valleys, the Q and Q' valleys of MoS$_2$ were lowered by $0.1$ eV.

The phonons that significantly couple at the Fermi level are : intravalley LA, TA,  LO, and A$_1$, as well as intervalley zone border acoustic phonons. 
As the coupling between valleys of opposite spins is at least an order of magnitude smaller than same-spin couplings, it is neglected. 
On the one hand, standard free-carrier screening increases with doping, diminishing the strength of field-mediated, small-momenta (intravalley) couplings like Fröhlich and piezoelectric interactions.
On the other hand, the screening of intravalley coupling to the A$_1$ and LA modes can counter-intuitively decrease as a function of doping when both K and Q types of valleys are occupied. The couplings are then enhanced.

\begin{figure*}
    \centering
\includegraphics[width=0.49\linewidth]{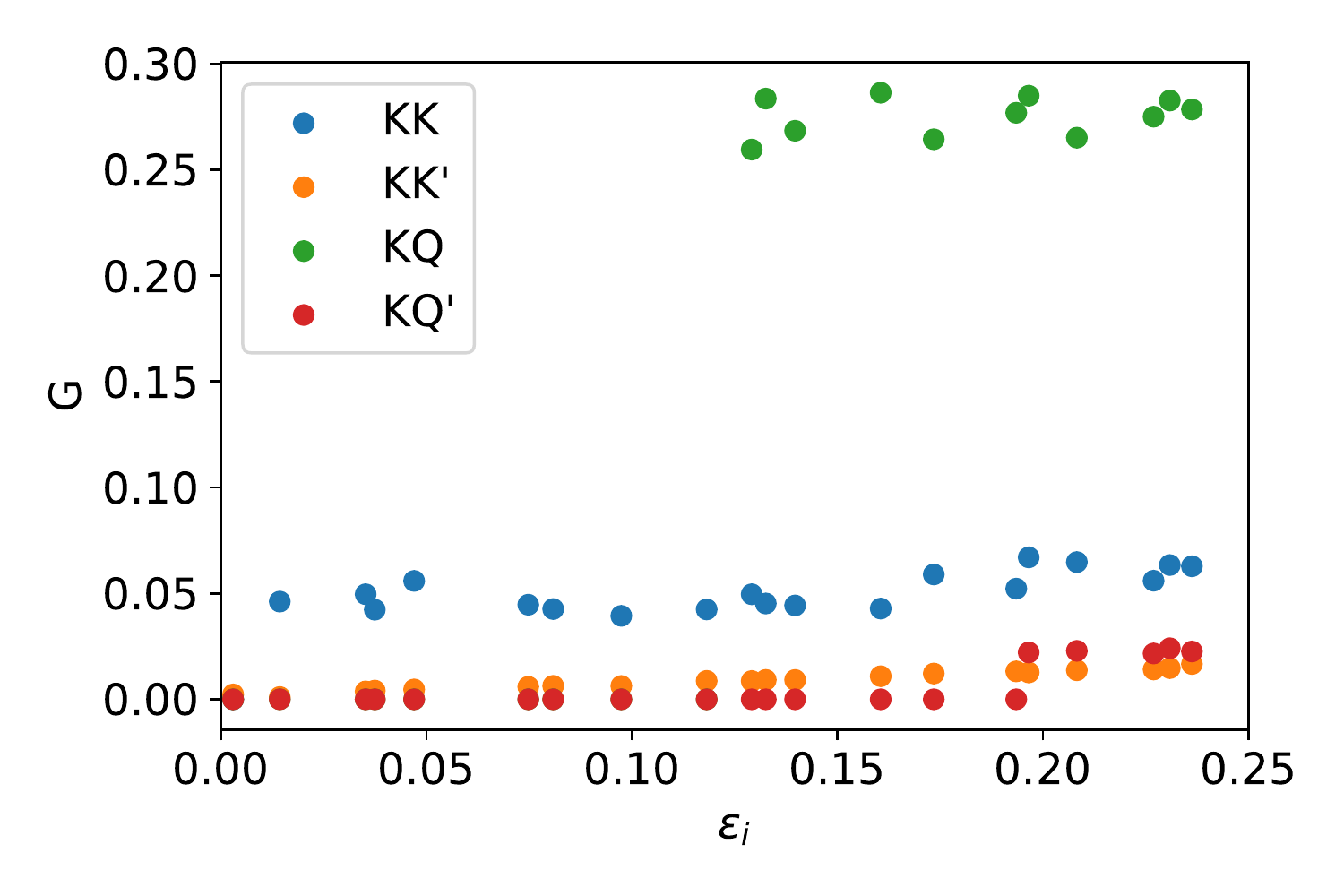}
\includegraphics[width=0.49\linewidth]{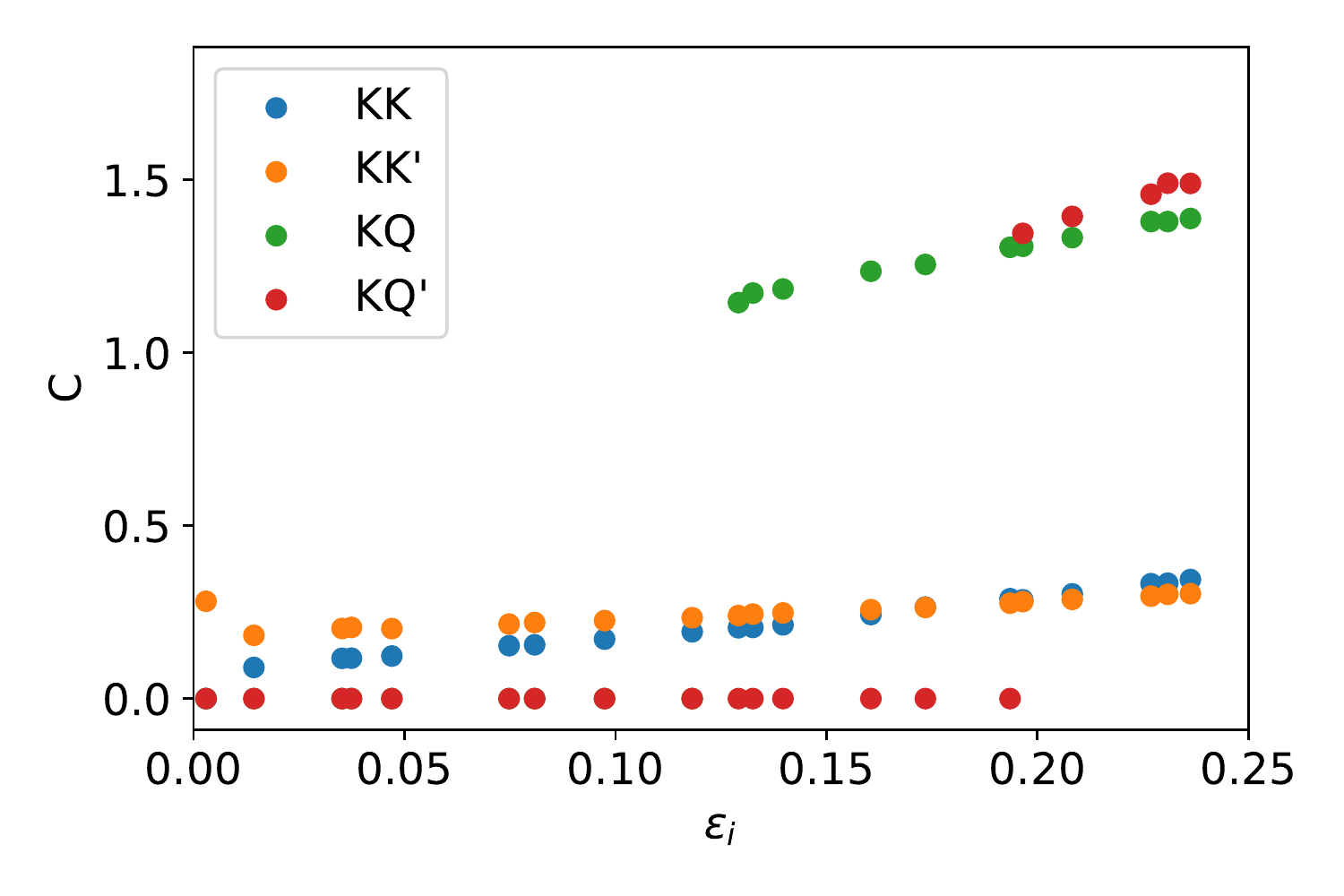}
    \caption{ Matrix elements of phonon-mediated attraction and Coulomb repulsion for all initial states $\mathbf{k}_i$ of energy $\varepsilon_i$ in the irreducible wedge of the Brillouin zone, showing negligible angular variations of the matrix elements. 
    }
    \label{fig:MoS2-GVvski}
\end{figure*}

We do not compute the phonon softening coming from the occupation of the Q valley. The phonon frequencies are those obtained with only the K valleys occupied.
It has been shown that the LA(M) phonon softens significantly when the corresponding electronic transitions are allowed, i.e. when the Fermi level crosses the Q valley.
According to anharmonic phonon calculations in multilayer MoS$_2$ \cite{Marini2023}, we can expect the LA mode to soften at M. The frequency decreases as a function of carrier concentration and potentially vanishes, leading to a charge density wave.
The computation of a realistic softening value is challenging and out of scope here. Our objective here is more the description of the nature of the solutions and the understanding of the mechanisms behind. 
We simply apply a $30\%$ softening of the LA(M) mode when Q is occupied. 

Eq. \ref{eq:matrix}, describes the couplings (phonon-mediated attraction or Coulomb repulsion) between different valleys. The final states are summed on the arrival valley $w$. In principle, one would average on the initial states at the Fermi level in $v$. 
Fig. \ref{fig:MoS2-GVvski} shows the matrix elements $G_{vw}$ and $C_{vw}$ for all initial states sampled in the irreducible wedge of the Brillouin zone. Plotted as a function of the Fermi-level, we obtain well defined lines of minimal widths. This indicates that those quantities are mostly dependent on the energy of the initial state and not so much on its momentum. This justifies the procedure of simply picking an initial state at the Fermi level rather than averaging over all states at the Fermi level.

\subsubsection{Screened Coulomb interaction}
\label{app:screened_coulomb}

To evaluate the dielectric function of the doped system, we simply sum the irreducible response functions associated to the neutral material $\chi^0_{\rm neut}$ and the added electrons in the conduction band $\chi^0_{\rm cond}$. The former is extracted from DFPT on the neutral material as in Ref. \cite{Sohier2021,Macheda2024}. The latter is computed from the band structure as:
\begin{align}
\chi^0_{\rm cond}(q) &= \sum_{n}  \int \frac{d^2\mathbf{k}}{(2\pi)^2} \frac{n^{\textrm{FD}}_{\varepsilon_{n\mathbf{k}}} - n^{\textrm{FD}}_{\varepsilon_{n\mathbf{k}+\mathbf{q}}}}{\varepsilon_{n\mathbf{k}} - \varepsilon_{n\mathbf{k}+\mathbf{q}}} 
|\braket{u_{n\mathbf{k}}|{u_{n\mathbf{k+q}}}}|^2
\end{align}
where $n^{\textrm{FD}}_{\varepsilon_{n\mathbf{k}}}$ is the Fermi-Dirac occupation for electronic state $\mathbf{k}$ of energy $\varepsilon_{n\mathbf{k}}$. The wavefunction overlaps are note computed. 
In general, it is $1$ for $\mathbf{q} \to 0$ (same electronic state), and decreases as $|\mathbf{q}|$ increases. It is expect to be much smaller than $1$ for intervalley transitions, e.g. when $\mathbf{q} \sim \mathbf{M}$, such that $\chi^0_{\rm cond}(q)$ becomes smaller for $|\mathbf{q}|$ larger than the size of the valleys. We assume the contribution from $\chi^0_{\rm cond}(q)$ to be negligible with respect to $\chi^0_{\rm neut}$, and use a simple complementary error function to model the wave-function overlaps, with a threshold roughly corresponding to the size of the valleys: $0.3$ bohr$^{-1}$.

\bibliography{biblioSCTMDs}
\bibliographystyle{myapsrev}

\end{document}